\documentclass[conference]{IEEEtran}
\IEEEoverridecommandlockouts
\usepackage{cite}
\usepackage{amsmath,amssymb,amsfonts}
\usepackage{graphicx}
\usepackage{textcomp}
\usepackage{xcolor}
\usepackage{listings}
\usepackage{physics}
\usepackage[linesnumbered,ruled,vlined]{algorithm2e}
\usepackage[noend]{algpseudocode}
\usepackage{multirow}
\usepackage{float}
\usepackage{tikz}
\usetikzlibrary{quantikz}
\usepackage{tabularx}
\usepackage{listings-rust}
\usepackage{caption}
\usepackage{subcaption}
\usepackage{flushend}

\def\BibTeX{{\rm B\kern-.05em{\sc i\kern-.025em b}\kern-.08em
    T\kern-.1667em\lower.7ex\hbox{E}\kern-.125emX}}

\DeclareMathOperator{\Det}{det}

\begin{document}
\title{\huge QudCom: Towards Quantum Compilation for Qudit Systems\\
}

\author{\IEEEauthorblockN{Daniel Volya and Prabhat Mishra}
\IEEEauthorblockA{\textit{Department of Computer \& Information Science \& Engineering} \\
\textit{University of Florida, Gainesville, Florida, USA}
}
}

\maketitle

\begin{abstract}
Qudit-based quantum computation offers unique advantages over qubit-based systems in terms of noise mitigation capabilities as well as algorithmic complexity improvements. However, the software ecosystem for multi-state quantum systems is severely limited. In this paper, we highlight a quantum workflow for describing and compiling qudit systems. We investigate the design and implementation of a quantum compiler for qudit systems. We also explore several key theoretical properties of qudit computing as well as efficient optimization techniques. Finally, we provide demonstrations using physical quantum computers as well as simulations of the proposed quantum toolchain. \end{abstract}




    

\begin{figure*}
    \centering
    \includegraphics[width=1\linewidth]{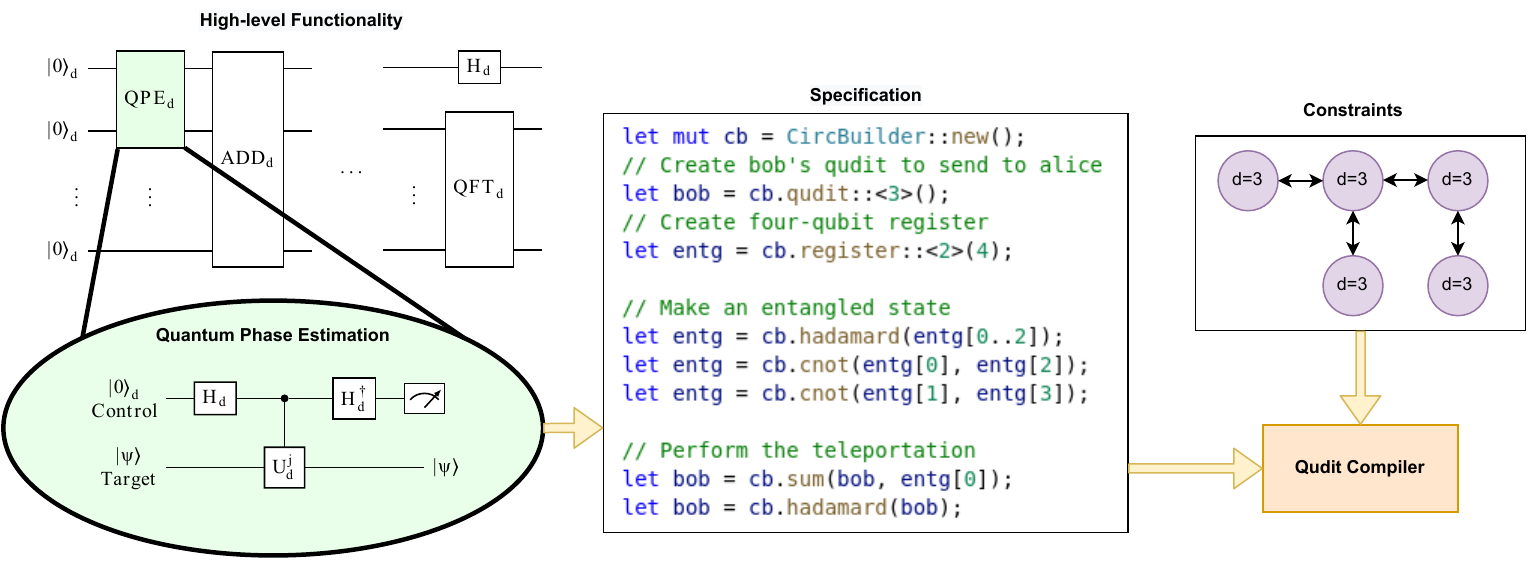}
    \caption{An example scenario for quantum compilation. The quantum phase estimation algorithm (shown as a circuit) is specified using a high-level language. This specification is compiled subject to constraints (qudit-connectivity).}
    \label{fig:flow}
\end{figure*}

\section{Introduction}
Quantum computing provides a promising alternative for solving problems that are hard for classical computers. A vast majority of quantum computation literature deals with qubits, a collection of two-state systems, and gates that yield arbitrary interactions between them. Under the assumption of arbitrary interactions, the computational space of the quantum computer scales as $2^N$, where $N$ is the number of qubits. The exponential growth in the state-space, and the ability to be in arbitrary superposition of these states, is one of the primary advantages over classical computation. However, one of the greatest challenges in engineering quantum computers is enabling interactions between the qubits, while also minimizing interactions with an environment and other sources of quantum and classical noise.

Recent efforts try to map quantum problems onto $d$-state (qudit) quantum computers \cite{gottesmanFaultTolerantQuantumComputation1999, wangQuditsHighDimensionalQuantum2020, hollandOptimalControlQuantum2020}. Early experimental methods have mapped problems to optimal control problem of multi-state systems, or qudits. Such a computational system scales in the order of $d^N$ where $N$ is the number of qudits. One of the primary objectives is that qudit systems will introduce greater noise tolerance as compared to strict qubit systems. This is in stark contrast to the leading methods of today -- using a collection of two-state units, or qubits \cite{IntroducingDesignAutomation,meuliSATbasedCNOTQuantum2018}. In addition to exploiting natural properties of a physical system for noise tolerance, qudit quantum computers could reduce space requirements. Specifically, quantum computing on higher-dimensional system can be more efficient than on qubits, and may even offer asymptotic computational improvements compared to qubit systems \cite{gokhaleAsymptoticImprovementsQuantum2019}. Moreover, there are entangled states on higher-dimensional systems that cannot be simulated by tensor products of pairwise entangled qubit states \cite{erhardExperimentalGHZEntanglement2018}.

\subsection{Contributions} 
Most commercial as well as academic quantum software operate purely in the qubit regime. Designing and simulating quantum circuits for qudit systems using these toolchains is often not feasible. In this paper, we seek to progress qudit research by introducing a quantum toolkit that operates on qudits. Rust is a language designed for performance and safety by using a type system with borrow checker rules. We utilize the inherent properties of the Rust compiler to ensure quantum mechanical properties remain satisfied, such as the no-cloning theorem or that dimensions of the intermediate qudits are compatible. In addition, we implement various state-of-the-art compilation and optimization techniques on qudit systems.  Specifically, this paper makes the following major contributions.
\begin{itemize}
    \item We propose a Rust-based quantum programming language to specify quantum algorithms for qudit systems.
    \item We implement fast cosine-sine decomposition heuristic as well as an optimal method based on Solovay-Kitaev algorithm for qudit systems.
    \item We enables evaluation of mixed quantum circuits (e.g. a qubit entangled with a qutrit.)
\end{itemize}

The remainder of this paper is organized as follows. Section~\ref{sec:background} provides relevant background and surveys related efforts. Section~\ref{sec:approach} describes our proposed quantum compilation framework for qudit systems. Section~\ref{sec:results} presents the experimental results. Finally, Section~\ref{sec:conclusion} concludes the paper.

\section{Background and Related Work}
\label{sec:background}
We first provide the background on quantum compilation as well as quantum circuits. Next, we survey related approaches to highlight the need for the proposed work.

\subsection{Quantum Compilation}

Quantum algorithms are typically described in a high-level language, such as in plain English or in pseudo-code, and then written in terms of a quantum circuit: a series of quantum gates applied to (traditionally) qubits in a sequential manner~\cite{volya2021quantum}. The quantum circuit ends with measurement, from which the results of the quantum algorithm are inferred via classical post processing~\cite{utt2023quantum}. Although imprecise, an analogy to classical computation is to define an instruction set which consists of a set of gates $\mathcal{G}$, and then compile to find a sequence of gates $g_1,\hdots,g_m$ drawn from the set $\mathcal{G}$ which approximates the original algorithm.

This transition from a high-level quantum algorithm to a quantum circuit is informally called ``quantum compilation", and consists of a series of steps which may depend on the particular quantum computer architecture. 
In general, the high-level algorithm and corresponding quantum circuit are invariant to the underlying assumptions of an architecture or technology. Specifically, after a quantum compiler outputs the instructions, a real-time controller device defines the actual interaction with a quantum computer, such as by coordinating many control devices that operate at the time-regime of quantum fidelity to send microwave pulses. 
Compiler optimizations can reduce noise and increase computation speed.  While long computations on near-term quantum computers may  be prohibitive due to constraints in circuit-depth, such computations can be performed by designing noise tolerant high-level algorithms and optimizing circuit translation with respect to noise.

\vspace{0.05in}
\noindent \textbf{Example 1}: Consider Figure~\ref{fig:flow}, where a quantum algorithm is specified utilizing high-level algorithmic blocks. Within each block is an underlying implementation of the algorithm's functionality. In this case, the example algorithm is the quantum phase estimation algorithm generalized to a d-dimensional qudits, where the underlying implementation is also represented as a quantum circuit. To specify arbitrary quantum circuits as well as to call higher-levels blocks for qudit systems, a Rust-based quantum programming language (RuQu) is used. The corresponding specification needs to be compiled to a set of basic logic gates based on device-specific constraints. $\blacksquare$

\subsection{Quantum Circuits}

Quantum computation involves the evaluation of the quantum circuit produced by the quantum compiler using a quantum computer. In this section, we briefly describe quantum computing using qubits as well as qudits.

\vspace{0.05in}
\noindent \textit{\underline{Qubit}}: A qubit is the quantum analogue of a classical bit. Namely, a qubit state $\ket{\psi}$ is defined as a linear combination of two basis states $\ket{0}$ and $\ket{1}$: $\ket{\psi} = \alpha \ket{0} + \beta \ket{1}$ where $|\alpha|^2 + |\beta|^2 = 1$. For $N$-qubits, the basis is a tensor product of individuals' basis. Hence the total size is $2^N$. 

\vspace{0.05in}
\noindent \textit{\underline{Qudit}}: A qudit is a linear combination of $d$-basis states. For example, a qutrit ($d=3$) state is described as: $\alpha \ket{0} + \beta \ket{1} + \gamma \ket{2}$ where $|\alpha|^2 + |\beta|^2 + |\gamma|^2 = 1$. Similarly a a $d$-dimensional qudit can be described as: $\alpha_1 \ket{0} + \alpha_2 \ket{0} + ... + \alpha_n \ket{d}$, where $|\alpha_1|^2 + |\alpha_2|^2 + ... + |\alpha_d|^2 = 1$. The tensor product of $N$ qudits, results in a total size of $d^N$. 

\vspace{0.05in}
\noindent \textit{\underline{Qubit Gates}}: A qubit quantum gate $U$ transforms a qubit from one state to another state. The operator is unitary, and can be described as a unitary matrix. In general, the group of $2 \times 2$ unitary matrices is the unitary group $U(2)$. A standard method for optimizing qubit gates is to use the decomposition
    $U = e^{i\gamma} e^{i (\lambda_0 \sigma_x + \lambda_1 \sigma_y + \lambda_2 \sigma_z)}$,
where $\sigma_x$, $\sigma_y$, and $\sigma_z$ are the Pauli matrices. The decomposition consists of an arbitrary phase factor from the $U(1)$ group, multiplied by an element of the special unitary group $SU(2)$ which is generated by the $su(2)$ Lie algebra.  The $SU(2)$ group is isomorphic to the group of quaternions of norm 1, and hence can be viewed as a representation for 3-dimensional rotations. Qubit compilers exploit this similarity, and find approximations for an algorithm by solving for Euler angles that achieve a desired rotation in 3-dimensional space. 

\vspace{0.05in}
\noindent \textit{\underline{Qudit Gates}}: In the general qudit case, the group of possible transformations are represented by $d \times d$ unitary matrices which form the $U(d)$ unitary group. Similar to the qubit case, an arbitrary $d \times d$ matrix $U$ can be decomposed into a $U(1)$ phase multiplied by an $SU(d)$ operator
$U = e^{i\gamma} e^{i \vec{\lambda} \cdot \vec{\sigma}}$, 
where $\vec{\sigma}$ represents the $d^2 - 1$ basis elements that span the $su(d)$ Lie algebra. However, unlike the qubit case, there is no obvious analogy with 3-dimensional rotations and hence requires alternative strategies. To the best of our knowledge, there are no existing efforts on effective quantum compilation for qudit systems. Instead one would have to make use of a qubit compiler and manually map qudit operations onto several qubits. Such an indirect and manual approach is not feasible for real-world applications. Most importantly, it loses some of the inherent advantages due to transformation via qubits.

\vspace{0.05in}
\noindent \textit{\underline{Qudit Clifford Group}}: An important group of gates used for generating entangled states and performing quantum error correction is the Clifford group. We briefly review the generalization of the Pauli group and the Clifford group to qudits. The Pauli group for qubits is defined via the Pauli operators and identity: $\{I, \sigma_x, \sigma_y, \sigma_z\}$. To generalize the Pauli group to qudits, the standard approach is define define operators
\begin{equation}
    X = \sum_{j=0}^{d-1} \ket{j}\bra{j+1}, \quad Z = \sum_{j=0}^{d-1} \omega^j \ket{j}\bra{j}
\end{equation}
where $\omega = e^{2\pi i/d}$ is the root of unity. The operators generalize $\sigma_x$ and $\sigma_z$ in the qubit case, and respect $X^d = Z^d = I$. For $n$-qudits, the operation acting on the $i$-th qubit is denoted with a subscript, e.g. $X_i$ and $Z_i$. A Pauli product is then defined as
\begin{equation}
    \omega^\lambda X^{\vec{x}} Z^{\vec{z}} = \omega^\lambda X_{0}^{x_0} Z_{0}^{z_0} \otimes X_{1}^{x_1} Z_{1}^{z_1} \otimes \dots \otimes X_{n}^{x_n} Z_{n}^{z_n}
\end{equation}
where $\lambda$ is part of $\mathbb{R}_d$ and $\vec{x}$ and $\vec{z}$ are tuples of length $n$ in $\mathbb{Z}_d^n$ -- each element an integer mod $d$. We can encode a single qudit Pauli element by decomposing it into two parts: $x \in \mathbb{R}_D$ and $y \in \mathbb{R}_D$. For example, in the qubit case, we have
$$I \Rightarrow (0, 0), \; X \Rightarrow (1, 0), \; Y \Rightarrow (1, 1), \; Z \Rightarrow  (0, 1)$$

In this encoding, multiplication of two Pauli elements $p_1$ and $p_2$ is additional mod $D$ in the encoding representation.

$$p_1p_2 \Rightarrow (x_{1,2}, y_{1,2}) = (x_1, z_1) + (x_2, z_2) \text{mod} D$$

For a fixed $n$, the Pauli group $\mathcal{P}_n$ is defined by all possible Pauli products. As an example, the Pauli group of a  single qutrit ($n=1$, $d=3$) is given by $\omega^\lambda X^iZ^j = \omega^\lambda[I, Z, ZZ, XZ, XZZ, XXZ, XXZZ]$ for $i, j \in \mathbb{R}_3$, and will have a total of 21-elements due to each 3 possible options for $\lambda$.

A Clifford operation $C$ acts on an element $p_1 \in \mathcal{P}_n$ such that under conjugation it returns another member of the Pauli group: $Cp_1C^\dagger = p_2$. Trivially, all Pauli products are Clifford operations. Any Clifford gate can be summarized by a tableau of its action on the generators $X$ and $Z$. Due to the group properties of the intersection of gates that stabilize a state \cite{gottesmanStabilizerCodesQuantum1997b}, $U\ket{\psi} = 1\ket{\psi}$, and Clifford gates, it is possible to simulate such gates efficiently on a classical computer \cite{aaronsonImprovedSimulationStabilizer2004c}.


\begin{figure*}[t]
  \includegraphics[width=1\linewidth]{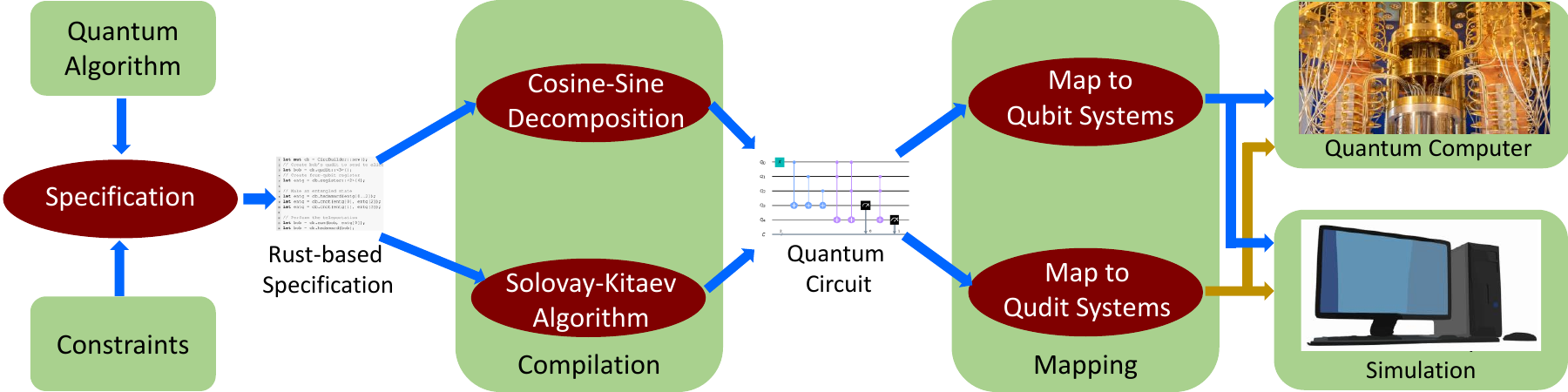}
  \centering
      \vspace{-0.2in}
  \caption{An overview of our proposed quantum compilation framework for qudit systems. The framework consists of three major tasks: specification, compilation and mapping/evaluation. The first task captures a quantum algorithm using Rust-based specification. The second task performs quantum compilation using either fast cosine-sine decomposition heuristic or slow but optimal Solovay-Kitaev alorithm. The third task performs mapping of the quantum circuit to enable evaluation using either a physical quantum computer or simulation of quantum systems.}
  \label{fig:compiler}
  \vspace{-0.2in}
\end{figure*} 

\subsection{Related Work}

Quantum compiler is a computer program that takes an arbitrary unitary matrix as input and returns the decomposition as a sequence of elementary operations. Due to the common assumption that a quantum computer operates on qubits, the input unitary matrix is often restricted to some power of two. Qubiter \cite{tucciRudimentaryQuantumCompiler1999} is one of the first quantum compilers to be proposed, it employs Quantum Bayesian Nets \cite{tucciQuantumBayesianNets1995} to model quantum systems graphically and utilizes the cosine-sine decomposition for compilation. To this day, Qubiter remains as one of the few quantum compilers that are general purpose -- there are no restrictions to the kinds of qubit unitary matrices to serve as input. However, when handling large qubit systems, or when required to work with a particular quantum technology with noise models, general purpose quantum compilers face unpractical computation times. For this reason, many state of the art quantum compilers restrict inputs to single-qubit and two-qubit gates. Most present-day quantum compilers such as the ones employed in Qiskit \cite{Qiskit} use the KAK decomposition \cite{tucciIntroductionCartanKAK2005} to exactly decompose a two-qubit operation. However, there is a growing research area to bring quantum compilers closer to the qudit systems. A theoretical investigation of quantum Shannon decomposition for qudit systems has been performed \cite{diSynthesisMultivaluedQuantum2013a}.

While there are many successful quantum compilers for qubit systems, to the best of our knowledge, there are no toolchains for qudit systems. Our proposed work utilizes the theoretical foundation of compilation and mapping algorithms to enable quantum compilation for qudit systems.

\section{Quantum Compilation for Qudit Systems}
\label{sec:approach}

Figure~\ref{fig:compiler} shows an overview of our proposed quantum compilation framework for qudit systems that consists of three major tasks. The first task enables specification of quantum algorithms using Rust-based programming language (Section~\ref{sec:specify}). The second task transforms a given specification to a quantum circuit (qudit) representation using either a fast cosine-sine decompositon heuristic \cite{tucciRudimentaryQuantumCompiler1999} (Section~\ref{sec:CSD}) or an optimal Solovay-Kitaev algoritm \cite{dawsonSolovayKitaevAlgorithm2005} (Section~\ref{sec:SK}). The third task performs mapping to enable evaluation using a physical quantum computer (Section~\ref{sec:quantum}) or simulation of qudit systems using classical computer (Section~\ref{sec:simulation}).
The remainder of this section is organized as follows. We first formulate the quantum compilation (synthesis) problem. Next, we describe each of the steps in our proposed framework in detail.

\subsection{Problem Formulation}

We formulate the task of decomposing an arbitrary unitary gate in terms of algebraic properties. Namely, a quantum gate $U$ can be represented as a unitary matrix, which belongs to the Lie group $U(d)$. By ignoring a phase factor $e^{i\theta} = \Det U$, then $U$ will belong to the $SU(d)$ Lie group.
Each special unitary matrix $U$ can then be written in the form:
\vspace{-0.05in}
\begin{equation}
    U = e^{-iL}
    \vspace{-0.05in}
\end{equation}
where $L$ is now a matrix from the $su(d)$ Lie algebra. $L$ can be expressed as a linear combination
  \vspace{-0.05in}
\begin{equation}
    L = \vec{L} \cdot \vec{\Lambda} = \sum_{j=1}^{d^2 - 1} L_j \Lambda_j, \qquad L_j \in \mathbb{R}
      \vspace{-0.05in}
\end{equation}
with the set $\vec{\Lambda}$ forming the basis for the $su(d)$ Lie algebra. For $d=2$, $\vec{\Lambda}$ are typically the Pauli matrices. For $d=3$, $\vec{\Lambda}$ are typically the Gell-mann matrices. For a general $d$ there are variety of possible representations for $\vec{\Lambda}$. These can be effectively represented using  generalized Gell-Mann matrices. The basis elements have a specific relation which allows defining an anti-symmetric cross product,
    \vspace{-0.05in}
\begin{equation}
    (\vec{A} \otimes \vec{B})_j = f_{jkl} A_k B_l = -(\vec{B} \otimes \vec{A})_j
        \vspace{-0.05in}
\end{equation}
and a symmetric dot product,
    \vspace{-0.05in}
\begin{equation}
    (\vec{A} \odot \vec{B})_j = d_{jkl} A_k B_l = +(\vec{B} \odot \vec{A})_j,
        \vspace{-0.05in}
\end{equation}
where $f_{jkl}$ is defined by the structure constants for the algebra, and $d_{jkl}$ is a symmetric tensor of coefficients.

Writing (anti-) commutation relations with cross and dot products, a compact equation for multiplication is derived:
    \vspace{-0.05in}
\begin{equation}
    (\vec{A} \cdot \vec{\Lambda})(\vec{B} \cdot \vec{\Lambda}) = \frac{2}{n} \vec{A}\cdot\vec{B} I_n + (\vec{A}\odot\vec{B} + i \vec{A} \otimes \vec{B}) \cdot \vec{\Lambda}.
        \vspace{-0.05in}
\end{equation}
Importantly, the equation lets us write any expression quadratic in the basis elements as a linear combination of them. 
A technique for quantum synthesis is to take a quantum gate $U = e^{iL}$, as well as basis gates $g_1 = e^{iG_1}$, $g_2 = e^{iG_2}$, $\hdots$, $g_n = e^{iG_n}$ and write $e^{iL} = e^{iG_a}e^{iG_{b}}\hdots$, and use (anti-) commutation relations to find $a,b,\hdots$ that specify which basis gate to use.

One of the key functionalities of a quantum compiler is to take an arbitrary gate $U$ and find a product of gates $g_1\hdots g_n$ (e.g. the universal set $g \in \{\text{SUM}, H, T\}$ \cite{luoUniversalQuantumComputation2014}) which approximates $U$. A naive, brute force approach is to find all possible permutations of the universal set and return the permutation that is closest to $U$. Such a construction would yield k-dimensional tree where $k$ is the size of the universal set. In general, there are two synthesis approaches: perform unitary decomposition using linear algebra techniques or empirical search-based techniques. The quality of a given approach is evaluated by the produced circuit depth and the difference between the solution and the original unitary. The Cosine-Sine decomposition (CSD) is the state-of-the-art linear algebra technique that provides a bound on circuit depth. On the other hand, the Solovay-Kitaev (SK) algorithm is a search technique that guarantees finding a gate sequence to approximate $U$ is possible, and is efficient to do so in $O(log^c(\frac{n}{\epsilon}))$, where $\epsilon >0$. The algorithm first starts by selecting an arbitrary gate $\tilde{U}$ that moves $U$ closer to the identity. Afterwards, the commutation relations in the $su(d)$ Lie algebra place a guarantee that the subsequent approximations of $\tilde{U}$ will converge fast. In practice, the algorithm is slow due to the arithmetic in $su(d)$, and can be numerically unstable. Despite the algorithm providing a theoretical guarantee to a solution, alternative methods such as CSD are used for gate synthesis. The next section describes how to synergistically combine SK and CSD algorithms for compiling quantum circuits.

 \begin{figure}[htp]
    \centering
    \vspace{-0.2in}
    \begin{subfigure}{0.35\columnwidth}
        \centering
        \includegraphics[width=\linewidth]{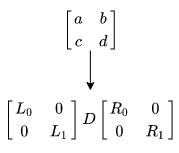}
        \caption{The Cosine-Sine decomposition} 
        \label{fig:cs-decomp}
    \end{subfigure}
    \hfill
    \begin{subfigure}{0.6\columnwidth}
        \centering
        \includegraphics[width=\linewidth]{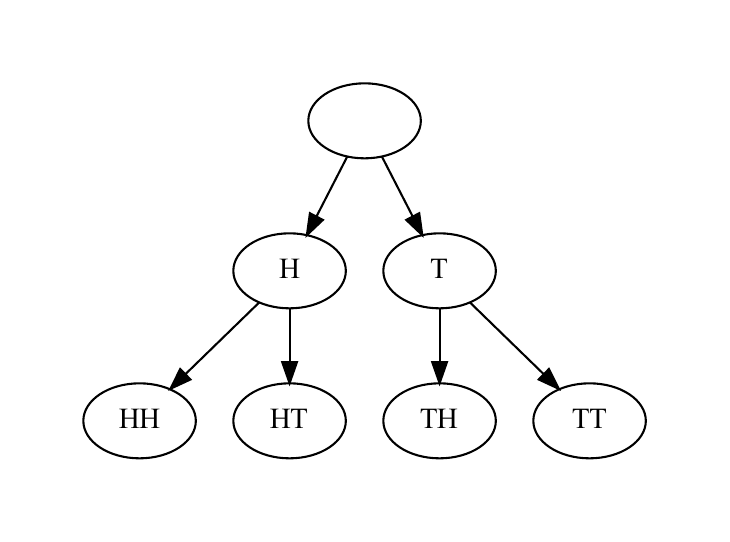}
        \vspace{-0.4in}
        \caption{Search for a sequence of gates}
        \label{fig:linear-search}
    \end{subfigure}
    \caption{Quantum compilation has two primary approaches: (a) linear algebra techniques or (b) empirical search.}
    \label{fig:synthesis}
\end{figure}

\subsection{Specification of Quantum Algorithms}
\label{sec:specify}

An important requirement for any toolchain or design automation task, including compilation, is the ability to capture the intent in an executable specification. In other words, an automated tool should be able to analyze the specification and derive the required executable models~\cite{volya2022modeling}. Specifically for quantum compilation, we need to capture the intent (quantum algorithm) as well as constraints (quantum architecture) using a specification language. There are promising efforts in utilizing Python based specification of quantum algorithms~\cite{Qiskit}, where the designer has to explicitly describe quantum constraints. In this paper, we use Rust-based specification since it inherently allows various checks suitable for quantum circuits that are detailed in the following sections.

\subsubsection{Qudit-based Quantum Programming Language}

Our Rust-based Quantum (RuQu) programming language which utilizes Rust's borrow checker to ensure the quantum program does not break the no-cloning theorem. Moreover, constant-generics are used to match the dimensionality of qudits during compile time. The primary mechanism to interface with RuQu is through the `CircBuilder' which provides convenient functions for constructing qudits and applying operations. 

\begin{lstlisting}[language=Rust, style=boxed, caption={Constructing a quantum circuit}, label={lst:qc}]
let mut cb = CircBuilder::new();
// Define a qutrit
let qutrit = cb.qudit::<3>();
// Define a circuit
let qutrit = cb.hadamard(qutrit);
// Measure
let (qutrit, m) = cb.measure(qutrit);
\end{lstlisting}

Listing~\ref{lst:qc} provides an illustrative example specification. Line 1 initializes a CircBuilder which keeps track of global information, such as the total number of qudits. Line 2 declares a single qudit with dimension 3, a qutrit. The dimension of the qudit must be known at compile-time, in order to satisfy constant generics and provide compile-time validation. Line 3 starts the circuit, which specifies that the generalized Hadamard is applied to the qutrit. Followed by a measurement as shown in Line 4.

Additionally, due to Rust's borrow checker, the no-clone theorem can be checked at compile time. For example, Listing~\ref{lst:clone} will fail since the quantum program requires copying a quantum state. Line 1 defines a function that will do something with the qudit, and is accessed on line 6. In line 7, we try to reuse the qudit, but it fails since the qudit was previously moved to the function. Otherwise, the state of the qudit would need to be cloned in order for both the function and the Hadamard operation to use it.

\begin{lstlisting}[language=Rust, style=boxed, caption={Failing no-clone theorem}, label={lst:clone}]
fn hold_my_qudit<Q>(_: Q) {}

let mut cb = CircBuilder::new();
let q = cb.qudit::<5>();
// Do something with the qudit
hold_my_qudit(q);
let q = cb.hadamard(q); // this fails
\end{lstlisting}   

Listing~\ref{lst:tele} shows a complex example of the quantum teleportation protocol using qubits.

\begin{lstlisting}[language=Rust, style=boxed, caption={Constructing the quantum teleportation protocol}, label={lst:tele}]
let mut cb = CircBuilder::new();
// Create bob's qudit to send to alice
let bob = cb.qudit::<3>();
// Create four-qubit register
let entg = cb.register::<2>(4);

// Make an entangled state
let entg = cb.hadamard(entg[0..2]);
let entg = cb.cnot(entg[0], entg[2]);
let entg = cb.cnot(entg[1], entg[3]);

// Perform the teleportation
let bob = cb.sum(bob, entg[0]);
let bob = cb.hadamard(bob);
\end{lstlisting}

\subsubsection{Building Blocks of Quantum Algorithms}

Quantum computers are promising for a wide variety of applications, such as for finding energy eigenstates of complex molecules or for factoring prime numbers. Certain algorithmic functionality, which is common among various applications, can be uniformly represented as fundamental building blocks. Figure~\ref{fig:flow} shows a high-level algorithm specified in terms of these fundamental blocks, such as the Quantum Phase Estimation (QPE) algorithm.

In the era of near and present-term quantum computers, where the number of qudits is relatively small (about 100), hand-writing quantum algorithms is feasible for experts. However, at a certain point in qudit count, manually creating, testing, and maintaining a quantum algorithm becomes impractical. As quantum computers scale to thousands of qudits, manually writing code for every qudit and every functionality will be impractical. In the attempt to increase the user experience in writing quantum algorithms, our framework includes common fundamental algorithmic building blocks. The user may conveniently specify various blocks that they wish to include in their quantum algorithms, without the worry of manually writing these blocks or testing the functionality for correctness. Listing~\ref{lst:blocks} shows an example circuit, where instead of defining gate-by-gate interactions between qutrits, functions are called to facilitate higher-level quantum algorithms.

\begin{lstlisting}[language=Rust, style=boxed, caption={Constructing a quantum circuit using blocks}, label={lst:blocks}]
let mut cb = CircBuilder::new();
// Declare 100 qutrits
let qutrits = cb.register::<3>(6);
// Perform generalized-X operation
// 2 qutrits
let qutrits = cb.X(qutrits[0..2]);
// Quantum phase estimation
// on all 5 qutrits
let qutrits = cb.QPE(qutrits[0..5]);
// Quantum Fourier transform
let qutrits = cb.QFT(qutrits[0..5]);
// Measure
let (_, m) = cb.measure(qutrits);
\end{lstlisting}

\subsubsection{Mixed-Dimensional Qudit Registers}

Presently, most implementations to realize entangling quantum gates are primarily done using only one and two-qubit quantum operations. Important multi-qubit gates that find themselves in quantum error correction and Shor's algorithms, such as the three-qubit quantum Toffoli (CCNOT) and Fredkin (CSWAP) gates, require a theoretical minimum of five two-qubit gates. In the pursuit of experimentally realizing systems with better coherence, lower error rates and faster quantum gate operations, it is necessary to develop strategies to remedy the cost of implementing important quantum gates. Recent research efforts attempt to exploit natural qudit properties in quantum systems to simplify the implementations of many-qubit gates, such as by coupling qubits to a qutrit.

RuQu enables researchers to describe operations utilizing a mixed-dimensional register. Listing~\ref{lst:ccnot} provides an example implementation of a CCNOT operation performed using two qubits and qutrit  \cite{baekkegaardRealizationEfficientQuantum2019a}.

\begin{lstlisting}[language=Rust, style=boxed, caption={A CCNOT using a qubit and qutrit as control}, label={lst:ccnot}]
let mut cb = CircBuilder::new();
// Declare two qubits
let qubits = cb.register::<2>(2);
// Declare a qutrit
let qutrit = cb.qudit::<3>();
// Perform CCNOT gate
let reg = cb.ccnot(
    qubit[0],
    qutrit,
    qubit[1]);
// Measure
let (_, m) = cb.measure(reg);
\end{lstlisting}

\subsection{Cosine-Sine Decomposition (CSD) Algorithm}
\label{sec:CSD}

One of the main objectives of the compiler is to rewrite any arbitrary unitary, or quantum gate, to an approximately equivalent quantum circuit which is constructed only out of a finite set of gates. Algorithm~\ref{alg:CSD} shows major steps in Cosine-Sine Decomposition (CSD). Consider a qubit system which is acted on by a unitary matrix $U$ of size $2^N \times 2^N$. CSD will produce the following:
    \vspace{-0.1in}
\begin{equation}
    U = diag(L_1, L_2) \begin{pmatrix} C & -S \\ S & C \end{pmatrix} diag(R_1, R_2)
\end{equation}
where $L_1, L_2, R_1, R_2$ are block matrices of size $2^{N-1} \times 2^{N-1}$. $C$ and $S$ are $\cos\vec{\theta}$ and $\sin\vec{\theta}$ respectively, where $\vec{\theta}$ is given by the CSD. 

\begin{algorithm}
    \DontPrintSemicolon 
    \SetKwFunction{FMain}{CSD} 
    \SetKwProg{Fn}{Function}{:}{\KwRet} 
    \Fn{\FMain{Unitary Matrix $U$, dim $d$}}{
        $n \gets \log_d(\text{size}(U))$ \;
        $m_0 \gets d^n$ \;
        $r_0 \gets d^{n-1}$ \;
        \For{$j\gets1$ \KwTo $d-1$}{
            \While{Have Submatrices}{
                $CSD(U_i^{(j)}, m_{j-1}, r_{j-1})$ \;
                $m_j \gets m_{j-1} - r_0$ \;
                $r_j \gets r_{j-1}$ \;
            }
        } 
        Combine all Matrices\;
    }
    \caption{Cosine-Sine Decomposition\label{alg:CSD}}
    \end{algorithm}

For a general qudit system which is acted on by a unitary matrix $U$ of size $d^N \times d^N$, CSD is iteratively performed $2^{d-1} - 1$ times and will produce a series of block matrices and Cosine-Sine matrices.
In either case, the block matrices are viewed as quantum multiplexers. For example, if we have a system of qutrits then the block matrices will have a form of

\vspace{-0.1in}
{
\tiny
$$    \begin{pmatrix} B_0 & & \\ & B_1 & \\ & & B_2\end{pmatrix}
$$}
\vspace{-0.1in}

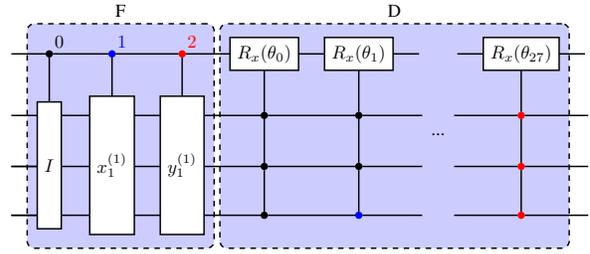
\begin{figure}[htp]
\centering
\vspace{-0.1in}
\scalebox{0.69}{
\begin{quantikz}
& \phase[black]{0} \gategroup[4,steps=3,style={dashed,rounded corners,fill=blue!20, inner xsep=2pt},background]{{\sc F}}&  \phase[blue]{1} & \phase[red]{2} & \gate{R_x(\theta_0)} \gategroup[4,steps=4,style={dashed,rounded corners,fill=blue!20, inner xsep=2pt},background]{{\sc D}} & \gate{R_x(\theta_1)} & \midstick[5,brackets=none]{...}\qw & \gate{R_x(\theta_{27})} &  \midstick[5,brackets=none]{...}\qw\\
& \gate[wires=3]{I} \vqw{-1} & \gate[wires=3]{x_1^{(1)}} \vqw{-1} & \gate[wires=3]{y_1^{(1)}} \vqw{-1} & \phase[black]{} \vqw{-1} & \phase[black]{} \vqw{-1} &\qw& \phase[red]{} \vqw{-1} & \qw \\
& \qw & & & \phase[black]{} \vqw{-1} & \phase[black]{} \vqw{-1}  & \qw& \phase[red]{} \vqw{-1} & \qw \\
& \qw & & & \phase[black]{} \vqw{-1} & \phase[blue]{} \vqw{-1} & \qw& \phase[red]{} \vqw{-1} & \qw
\end{quantikz}
}
\caption{An example circuit following the Cosine-Sine decomposition of $4$ qutrits.}
\label{fig:csd-qutrit}
\vspace{-0.1in}
\end{figure}

Equivalently, the quantum circuit will do the following: if the first qutrit is in state $\ket{0}$ apply $B_0$, if the first qutrit is in state $\ket{1}$ apply $B_1$, and if the first qutrit is in state $\ket{2}$ apply $B_2$. In general, a qutrit matrix will be decomposed into $U = ABC D EFG$, where $A,C,E,G$ are block matrices and $B, D, F$ are Cosine-Sine matrices. A visualization of the circuit is shown in Figure~\ref{fig:csd-qutrit}.

\vspace{0.05in}
\noindent \textbf{Example 2}:
Continuing from Example 1, the compilation of the example circuit using the CSD method yields: 

        \begin{quantikz}
            \lstick{$\ket{0}_d$\\Control} & \gate{H_d} & \ctrl{1} & \ctrl{1}& \ctrl{1}& \gate{H_d^\dagger} & \meter{} \\
            \lstick{$\ket{\psi}$\\Target} & \qw        & \gate{H_0^j} & \gate{H_1^j} &\targ{} & \qw  & \qw\rstick{$\ket{\psi}$}
        \end{quantikz}
        
The synthesis produces a circuit where the controlled operations are distributed into a product of controlled gate operations. $\blacksquare$


\subsection{Solovay-Kitaev (SK) Algorithm} 
\label{sec:SK}

The CSD algorithm discussed in the previous section is an efficient heuristics, but is not guaranteed to produce accurate results. In contrast, Solovay-Kitaev algorithm is optimal. In other words, it is guaranteed to be $\epsilon$-close to the expected output , given a desired error $\epsilon$. However, Solovay-Kitaev algorithm inherently utilizes tree-like structure, and although significantly limits the search space by exploiting algebraic properties, is slower than the CSD algorithm.

\begin{algorithm}
\SetAlgoLined
\DontPrintSemicolon
    \SetKwFunction{FMain}{Solovay-Kitaev}
    \SetKwProg{Fn}{Function}{:}{}
    \Fn{\FMain{Gate $U$, depth $n$}}{
        {\eIf{$ n == 0 $}
            { \textbf{return} Basic Approximation of $U$ }
            {
                $ U_{n-1} \longleftarrow \text{Solovay-Kitaev}(U, n-1)$\\
                $ V, W \longleftarrow \text{Approx-Decompose}(U, n-1)$\\
                $ V_{n-1} \longleftarrow \text{Solovay-Kitaev}(V, n-1)$\\
                $ W_{n-1} \longleftarrow \text{Solovay-Kitaev}(W, n-1)$\\
                \textbf{return} $U_n = V_{n-1}W_{n-1}V_{n-1}^\dagger W_{n-1}^\dagger U_{n-1}$
            }
        }
}
\textbf{End Function}
\caption{Solovay-Kitaev Algorithm}
\label{alg:SK}
\end{algorithm}

Algorithm~\ref{alg:SK} shows the major steps in Solovay-Kitaev algorithm. It uses the observation  that for an accuracy of $\epsilon > 0$, a sequence of gates that approximate the unitary can be generated in $O(\log^c (1/\epsilon))$. The underlying strategy is to start at an arbitrary approximation, which can be stored to a table ahead of time. Then, by utilizing the properties of $SU(d)$, keep applying transformation that drives the operation to a closer approximation until a circuit depth of $n$ is reached.

Although theoretically efficient, in practice the Solovay-Kitaev algorithm suffers from large runtimes primary due to the iterative search structure. For this reason, methods such as CSD are more common due to their faster runtime despite them not producing the most optimal solution. Moreover, methods such as CSD exploit stable and well studied matrix decompositions.

\vspace{0.05in}
\noindent \textbf{Example 3}:
Continuing from Example 1, the compilation of the example circuit using the Solovay-Kitaev algorithm yields:

\begin{quantikz}
    \lstick{$\ket{0}_d$\\Control} & \gate{H_d} & \ctrl{1} & \gate{H_d^\dagger} & \meter{} \\
    \lstick{$\ket{\psi}$\\Target} & \qw        & \gate{Z^j}  & \qw  & \qw\rstick{$\ket{\psi}$}
\end{quantikz}
        
Due the optimal search nature, SK-method figures out that the $U$ operation is simply the $Z$ operation as it is the ``closest" gate in the search tree. $\blacksquare$

\subsection{Synergistic Integration of SK and CSD Algorithms}

As discussed in the previous sections, the SK-algorithm guarantees an optimal solution, but may become impractical for larger quantum circuits. Alternatively, CSD method utilizes Cosine-Sine matrices to decompose an arbitrary unitary, and although efficient, the resulting decomposition is generally difficult to realistically implement. To combine their advantages, we implement the compilation in steps:
\begin{enumerate}
    \item Perform CSD on a unitary to obtain an intermediate-representation (IR) consisting of Cosine-Sine and block matrices.
    \item Perform SK on the IR to obtain a convenient representation consisting of $\{H, CNOT, T\}$
\end{enumerate}

By performing CSD first, we efficiently obtain a decomposition of a unitary matrix. This intermediate representation consists of Cosine-Sine and block matrices. To write these matrices in terms of a universal set of quantum gates, the SK algorithm is performed on the Cosine-Sine matrices and block matrices up to a defined error $\epsilon$. 

\subsection{Mapping to Qubit Systems}
\label{sec:quantum}

For certain problems, it may be easier to derive an algorithm in terms of d-dimensional qudits and then map to suitable e-dimension for a physical quantum computer. As an example, qubitization \cite{lowHamiltonianSimulationQubitization2019} is a common manual process to map from a higher-dimensional problem to qubits. In the following sections an automatic compilation strategy is introduced and evaluated. We evaluate the compiled circuits using physical qubit-based quantum computer. Let us illustrate the mapping procedure using an example. 

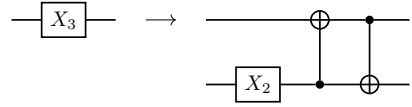
\begin{figure}[htp]
    \centering
    \vspace{-0.1in}
    \scalebox{0.8}{
    \begin{tikzcd}& \gate{X_3} & \qw & \arrow[r] & & & \qw & \targ{} & \ctrl{1} & \qw\\ & & & & & &\gate{X_2} & \ctrl{-1} & \targ{} & \qw \end{tikzcd}
    }
    \vspace{-0.1in}
    \caption{Example mapping of a qutrit circuit to a qubit circuit.}
    \label{fig:qudit_to_qubit}
    \vspace{-0.1in}
\end{figure}

We start with an example of mapping qutrit ($d = 3$) circuits to qubit circuits. Consider the $X_3$ gate, which can be represented in several ways on a qubit space. One possible matrix representation is as follows:

\begin{equation}
    \tilde{X_3} = X_3 \oplus I_1 = \begin{bmatrix} 0 & 0 & 1 & 0 \\ 1 & 0 & 0 & 0 \\ 0 & 1 & 0 & 0 \\ 0 & 0 & 0 & 1 \end{bmatrix}
\end{equation}
The corresponding circuit is shown in Figure~\ref{fig:qudit_to_qubit}. In addition, consider the qutrit Hadamard gate. Examples of representing the qutrit Hadamard matrix of two qubits include:

\begin{equation}
    \tilde{H}_a = \frac{1}{\sqrt{3}} \begin{bmatrix}
    1 & 1 & 1 & 0 \\
    1 & c & c^2 & 0 \\
    1 & c^2 & c & 0 \\
    0 & 0 & 0 & \sqrt{3}
\end{bmatrix}
\end{equation}
\begin{equation}
    \tilde{H}_b = \frac{1}{\sqrt{3}} \begin{bmatrix}
    \sqrt{3} & 0 & 0 & 0 \\
    0 & 1 & 1 & 1 \\
    0 & 1 & c & c^2 \\
    0 & 1 & c^2 & c \\
\end{bmatrix}, \quad c=e^{i\pi/3}
\end{equation}   

\begin{figure*}[h!]
    \begin{subfigure}[b]{0.5\textwidth}
    \includegraphics[width=1\columnwidth]{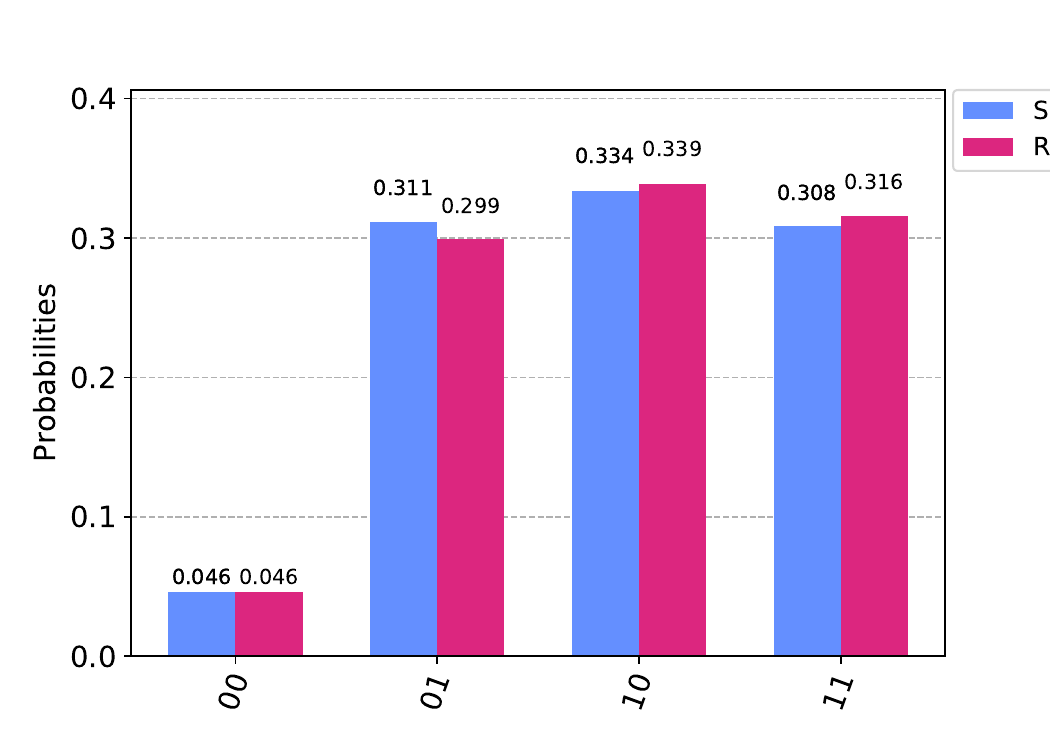}
    \vspace{-0.4in}
    \caption{$\tilde{H}_a$}
    \centering
    \end{subfigure}
     \begin{subfigure}[b]{0.5\textwidth}
    \includegraphics[width=1\columnwidth]{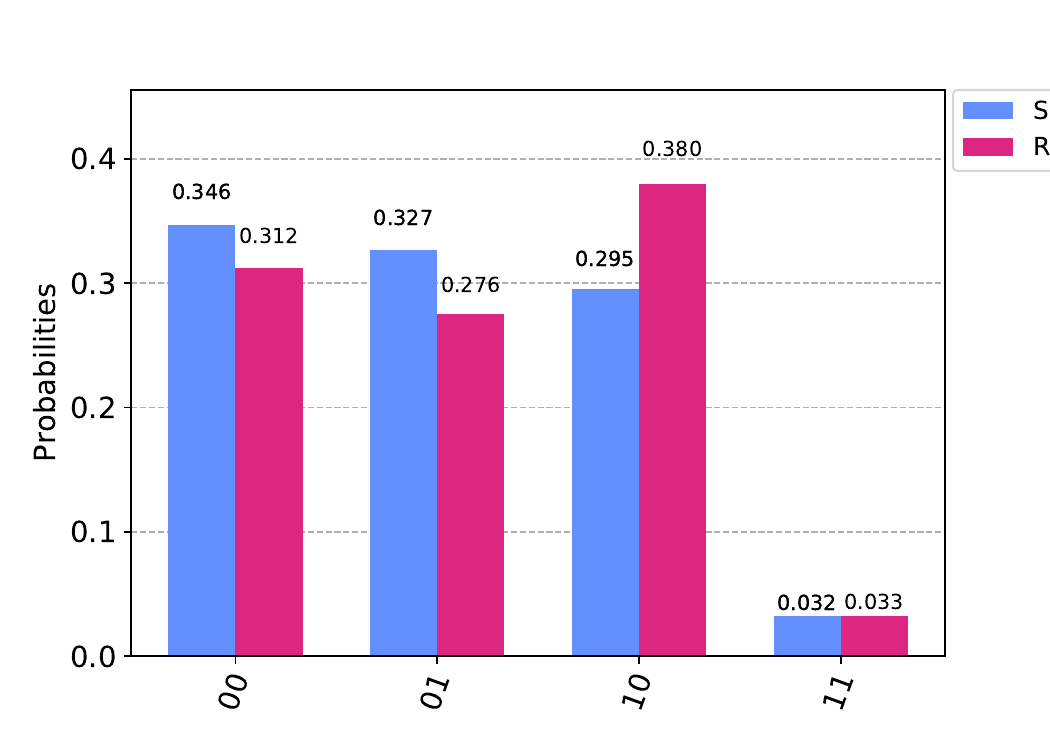}
    \vspace{-0.4in}
    \caption{$\tilde{H}_b$}
    \centering
    \end{subfigure}  
    \caption{The resulting state $\frac{1}{\sqrt{3}}(\ket{0} + \ket{1} + \ket{2})$. The resulting state for $\tilde{H}_a$ is closer to the expected result, suggesting that using the first qubit for the leakage state is better. S denotes the quantum computer simulator with a noise model, and R denotes ibmq\_melbourne. \label{fig:qutrithad}    }
    \vspace{-0.2in}
\end{figure*}  

\vspace{0.05in}
\noindent \textbf{Example 4}:
Compiling the qudit algorithm into a qubit algorithms first requires mapping the qudit algorithm into a qubit subspace.
        \begin{quantikz}
            \lstick{$\ket{0}_2^{\otimes n}$\\Control} & \gate{H_d}\qwbundle[alternate]{} & \ctrl{1}\qwbundle[alternate]{} & \gate{H_d^\dagger}\qwbundle[alternate]{} & \meter{}\qwbundle[alternate]{} \\
            \lstick{$\ket{\psi}$\\Target} & \qw        & \gate{U^j}  & \qw  & \qw\rstick{$\ket{\psi}$}
        \end{quantikz}
$\blacksquare$

\subsection{Mapping to Qudit Systems} 
\label{sec:simulation}
As the area of qudit algorithms and error correction grows, it is important that real-world experiments can be performed. In this section, we describe a compilation to map arbitrary qudit algorithms to other qudit ones.
In general, we may have a $d^n \times d^n$ unitary matrix $A$ that we wish to map to a system of $e$-dimensional qudits. To do this, find a $m$ such that $e^m \geq d^n$. Next let the difference be $l = e^m - d^n$. Define a new matrix $B$ of size $e^m \times e^m$ as:

\begin{equation}
B = A \oplus I_{l \times l} = \begin{pmatrix} A & 0 \\ 0 & I \end{pmatrix}
\end{equation}

Finally, we perform the CSD to write the matrix B in terms of $m$, $e$-dimensional qudits. In other words, the first step is to find the number of necessary qudits in order to completely capture the original operator $A$. In the case the space encoding the operator $A$ is larger than $A$, there may be many subspaces in which $A$ may reside. In real systems however, certain subspaces me be more preferable than others due to decoherence considerations.

\section{Experiments}
\label{sec:results}

In order to assess the performance of our qudit compiler, we look at suite of quantum gates and algorithms. The main goal is to demonstrate the various usage scenarios in compiling unitary gates. We compare the circuit sizes for qubit and qudit implementations. We use both a simulator and IBM-Q quantum computer to generate the results. Specifically, we used IBM-Q to show that our compiled output can be mapped to today's physical quantum computers. Since there are no available simulators for qudit systems, we have implemented a prototype simulator for qudit systems to evaluate our compiled qudit circuits.

\subsection{Evaluation using IBM-Q quantum computer}
We used IBM-Q qubit-based quantum computer to test mapping qudit systems to qubits. Due to decoherence times and limits in circuit depth of the quantum computers, only small-scale experiments were conducted. Namely, we focused on looking at single-gate operations which required a small qubit circuit depth.

\begin{table}[htp]
\begin{tabularx}{\columnwidth}{c | c | c }
    Operator & $d=2$ Gates & $d=3$ Gates\\
     \hline
    \begin{tikzcd} & \gate{X_d} & \qw \end{tikzcd} & H: 2, T: 4 & H: 4, T: 6 \\
    \begin{tikzcd} & \gate{Y_d} & \qw \end{tikzcd} & H: 2, T: 12 & H: 4, T: 14 \\
    \begin{tikzcd} & \gate{Z_d} & \qw \end{tikzcd} & H: 0, T: 4 & H: 0, T: 6 \\
     \hline
\end{tabularx}
\caption{A list of qudit-based quantum gates approximated by gates from the universal gate set $\{\text{SUM}, H, T\}$}
\label{tab:qudit-gates}
\end{table}

We observed the number of basis-gates required to implement other gates. Table~\ref{tab:qudit-gates} lists the number of single-qudit T and H gates required to implement the generalized Pauli Gates. Figure~\ref{fig:qutrithad} shows a qutrit Hadamard implemented on two qubits. The qutrit Hadamard gate takes H$\ket{00} \rightarrow \frac{1}{\sqrt{3}}(\ket{00} + \ket{01} + \ket{10})$. As shown in the figure, the transformation is not perfect and has some probability leakeage into the $\ket{00}$ or $\ket{11}$ state which lies outside of our qutrit space. In addition, within the qutrit space the probability is not uniform, showing that there is subspace noise. In this case, leakage was lower when encoding the qutrit into states $\ket{00}, \ket{01},$ and $\ket{10}$. However, the subspace noise was higher in this case.

\begin{figure}[H]
  \vspace{-0.2in}
  \includegraphics[width=0.6\linewidth]{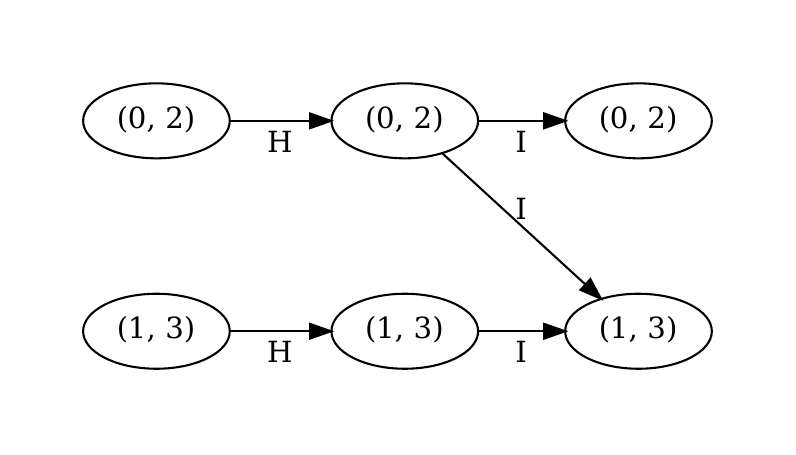}
  \centering
  \vspace{-0.2in}
  \caption {The directed acyclic graph produced for a quantum circuit consisting of a qubit and qutrit, followed by an application of a Hadamard gate to both qudits, and finished with a SUM operator}
  \label{fig:simulator-graph}
\end{figure}  

\subsection{Simulation of Qudit Systems}
\begin{figure}[h!]
    \centering
        \vspace{-0.2in}
    \begin{subfigure}{0.3\columnwidth}
        \centering
        \begin{tikzcd}[sep=tiny]& \gate{H_2} &  \ctrl{1} &  \meter{} \\ & \qw & \targ{} & \meter{} \end{tikzcd}    
        \caption{Example circuit}
    \end{subfigure}
    \hfill
    \begin{subfigure}{0.5\columnwidth}
        \centering
        \includegraphics[width=\linewidth]{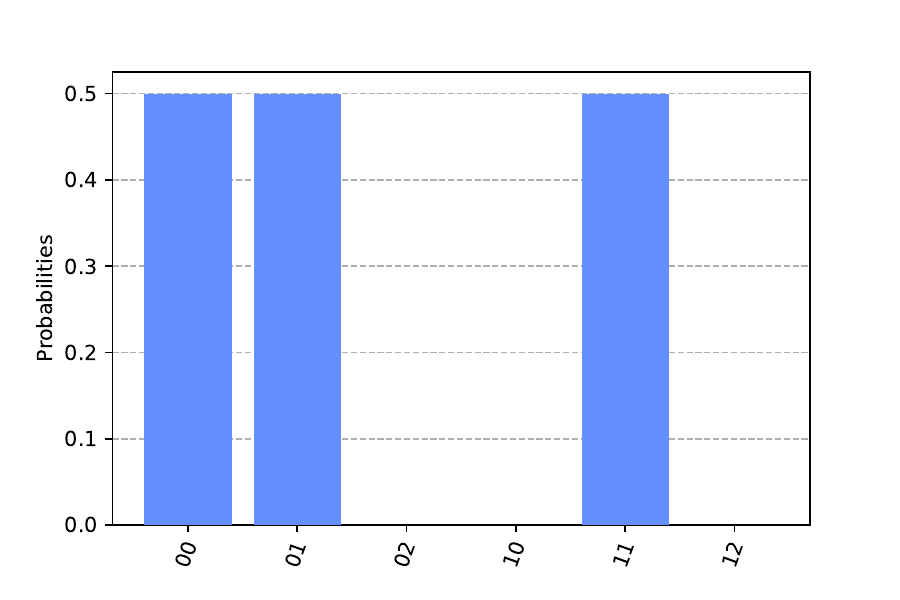}
        \vspace{-0.2in}
        \caption{Simulation results}
    \end{subfigure}
    \vspace{-0.1in}
    \caption{Qudit simulator results (b) simulating the circuit in (a) acting on a qubit and a qutrit via a SUM operator.}
    \label{fig:qudit-simulation}
    \vspace{-0.2in}
\end{figure}

Due to quantum circuit simulators focusing primarily on qubit systems, we developed a qudit circuit simulator. The simulator reads the quantum circuit as specified by RuQu and produces a directed acyclic graph with at most two edges at each vertex. Figure~\ref{fig:simulator-graph} shows an example directed acyclic graph that consists of a qubit and qutrit. Each edge represents an operation applied to the particular qudit. When two edges enter a vertex, the state of the qudits are added modulo the dimension of qudit represented by the vertex. From the example, the SUM gate is implemented using two edges that perform an identity operator to the state of the previous vertex and then adds the resulting states addition module three. The simulator is primarily used for validation of qudit circuits. Figure~\ref{fig:qudit-simulation} shows the results from running a circuit using our qudit simulator.

\subsection{Comparison of Compilation Strategies}
\begin{figure}[h]
    \centering
        \vspace{-0.1in}
    \includegraphics[width=\linewidth]{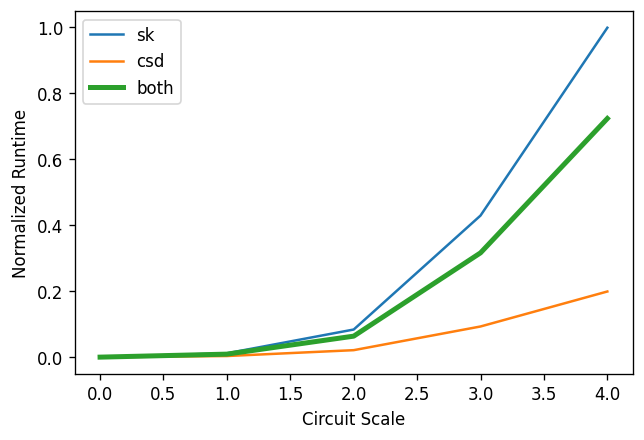}
         \vspace{-0.2in}
    \caption{Runtime of compiler by either (a) purely SK, (b) purely CSD, (c) or synergistic combination of SK and CSD}
    \label{fig:runtime}
    \vspace{-0.1in}
\end{figure}

By compiling an arbitrary unitary matrix by first utilizing CSD and then SK, we are able to minimize the runtime overhead while also providing a decomposition in terms of a given universal gate set. The minimization is achieved by noting that the intermediate-representation produced by CSD often has repeating entries, so the SK algorithm needs to be only run once. Figure~\ref{fig:runtime} showcases the runtime performance, where the synergistic combination achieves better performance than purely performing SK algorithm, while also maintaining the accuracy and desired output in terms of a universal gate set.

\section{Conclusion}
\label{sec:conclusion}
In this work, we have introduced multivalued elementary quantum gates. We have developed methods to compile an arbitrary quantum unitary into a sequence of qudit elementary gates. We have also explored methods for re-targeting qudit circuits to qubit circuits and vice-versa. Moreover, we have provided a quantum programming language using Rust to develop qudit-based quantum circuits that are validated during compile-time. We have evaluated the effectiveness of our proposed quantum compilation framework using physical quantum computer as well as prototype qudit simulator. With these sets of tools, we hope to bring more researchers to explore qudits as a suitable computational platform. Qudit-based quantum computation is a growing research area. In the area of qudit compilation and synthesis there are still plenty of open questions and potential improvements. The exploration into these topics may reveal additional advantages to qudit-based computers compared to qubit-based ones. Additionally, choosing an appropriate physical implementation of qudit systems and investigating the implementation of qudit gates will remain crucial to the development of qudit-based quantum computing. The future research needs to explore the compilation challenges in the context of state preparation~\cite{volyaStatePreparationQuantum2023a}, data compression~\cite{volyaQuantumDataCompression2023}, as well as secure~\cite{volya2023towards} and noise-resilient~\cite{volya2020special} quantum computing.

\bibliographystyle{IEEEtran}
\bibliography{compiler.bib}

\end{document}